# Nanoscale Modulation of Flat Bands via Controllable Charge-Density-Waves Defects in 4Hb-TaS$_2$

Wooin Yang[1,2], S. Karbasizadeh[4], Hoyeon Jeon[1], Saban Hus[1], Arthur P. Baddorf[1], Sai Mu[4], Tom Berlijn[1], Haidong Zhou[2], Wonhee Ko[2,3*], An-Ping Li[1,2*]

[1]*Center for Nanophase Materials Sciences, Oak Ridge National Laboratory, Oak Ridge, Tennessee 37831, USA*

[2]*Department of Physics and Astronomy, University of Tennessee, Knoxville, Tennessee 37996, USA*

[3]*Center for Advanced Materials and Manufacturing, University of Tennessee, 2641 Osprey Vista Way, Knoxville, Tennessee 37920, USA.*

[4] *Department of Physics and Astronomy, University of South Carolina, Columbia, 29208, USA*

* Email: wko@utk.edu, apli@ornl.gov

**Abstract**

Electron correlation is a main driver of exotic quantum phases and their interplay. The 4Hb-TaS$_2$ system, possessing intrinsic heterostructure of 1T- and 1H-TaS$_2$ monolayers, offers a unique opportunity to control electron correlation by distorting the atomic lattice or tuning interlayer coupling. Here, we investigated intrinsically deformed charge-density-waves (CDW) in the 1T layer of 4Hb-TaS$_2$ to elucidate and control their effects on flat bands using scanning tunneling microscopy and spectroscopy (STM/S) combined with first-principles calculations. We identified two types of CDW defects: Type 1 has structural distortion and locally suppressed flat bands, while Type 2 features an increased flat band filling factor of intact CDW structure. Density functional theory calculations indicate that a sulfur vacancy in the 1T layer distorts the CDW structure and gives rise to a Type 1, whereas a sulfur vacancy in the 1H layer reduces the interlayer charge transfer and lead to a Type 2. Furthermore, we demonstrated creating and erasing individual CDW defects via STM manipulation. Our findings provide a pathway to not only tune flat bands but also selectively manipulate the interaction between CDW, the atomic lattice, and interlayer coupling in strongly correlated systems with atomic precision.

Tuning the strength of electron correlation is a central topic in condensed matter physics, as it enables a deeper understanding of the origin of various strongly correlated states and offers access to exotic quantum phase. In particular, flat bands—where kinetic energy is quenched—significantly amplify electron correlations. By tuning flat band filling factor, we gain a controllable knob for adjusting the strength of electron correlation, potentially leading to the realization of exotic quantum phenomena such as unconventional superconductivity [1] and charge fractionalization [2,3].

In two-dimensional (2D) layered materials, weak interlayer coupling further confines electron motion, thereby increasing electron correlation [4,5]. Transition metal dichalcogenides (TMDC) are a family of 2D layered materials that provide a highly versatile template for controlling electron correlation due to the wide range of available chemical compositions [6-9]. Among them, $TaS_2$ stands out because its two polytypes exhibit distinct properties. In 1T-$TaS_2$, a $\sqrt{13}a \times \sqrt{13}a$ charge-density-wave (CDW) forms a superlattice of star-of-David (SoD) clusters, each consisting of 13 Ta atoms and 26 S atoms. The hybridized orbitals form half-filled flat bands near the Fermi level $E_F$, which results in a Mott insulator phase with incorporated Coulomb repulsion [10]. In contrast, 2H-$TaS_2$ remains metallic, with superconductivity emerging at a critical temperature $T_c$ = 0.8 K [11].

Recent studies have shown that these correlated states in $TaS_2$ can be further tuned by constructing 1T/1H heterostructures. In 1T/1H bilayer system, interactions between electrons in the 1T and 1H layers give rise to an artificial heavy fermion phase [9]. Meanwhile, in 4Hb-$TaS_2$, which has the intrinsic heterostructure of alternating 1T and 1H layers, interlayer coupling induces significant charge transfer from the 1T to the 1H layer, leading to the bulk crystal to adopt a heavily hole-doped Mott insulator phase [12]. This interlayer coupling also affects superconductivity in 4Hb-$TaS_2$, resulting in enhanced $T_c$ = 2.7 K [13], broken time-reversal symmetry [14] and non-trivial topology [13,15]. Moreover, previous studies have displayed that the modulation of the interlayer coupling occurs not only at extrinsically implanted atomic impurities [16,17] but also at particular SoD clusters of pristine 4Hb-$TaS_2$ [18]. However, the effects of native defects on the CDW-driven flat band remain to be examined.

Here we examine CDW defects, structurally or electronically perturbed SoD clusters in pristine 4Hb-$TaS_2$, and investigate their effects on flat bands using scanning tunneling microscopy/spectroscopy

(STM/S) and first-principles calculation. The identified CDW defects can be classified into two types, Type 1 and Type 2, based on their distinct interactions with SoD clusters and flat bands. Type 1 defects distort individual SoD clusters, leading to the suppression of flat band features, the emergence of defect states, and localized hole doping. In contrast, Type 2 defects induce electron doping accompanied by flat band splitting and the emergence of Kondo-like resonance. First-principles calculations indicate distinct origins for these defects; an atomically defective 1T layer hosts Type 1, while locally disturbed interlayer charge transfer causes Type 2 in a defect-free 1T layer. Furthermore, we demonstrated bi-directional manipulation of individual defects, enabling their creation and removal. Given that defect manipulation allows control over interlayer coupling [15], electronic doping levels [19], and magnetic moments [20], our results highlight that local tuning of correlated electron behaviors through CDW defects is a promising approach for controlling strongly correlated states at the atomic scale.

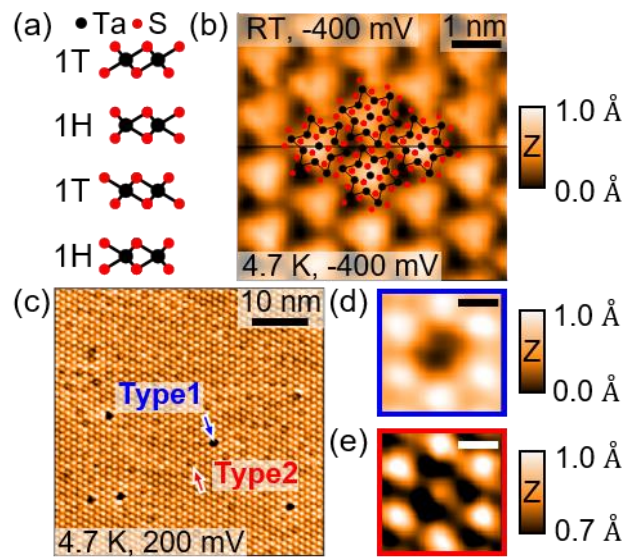

Fig. 1. (a) Atomic structure of 4Hb-TaS$_2$. Black circles represent Ta atoms, while red circles denote S atoms. (b) Two STM images of the 1T layer of 4Hb-TaS$_2$ measured at room temperature (upper panel) and 4.7 K (lower panel). Black and red circles are overlaid to indicate the positions of Ta atoms and top S atoms, respectively. (c) STM image of the 1T layer showing two types of CDW defects. A blue arrow points to a Type 1 defect, and a red arrow indicates a Type 2 defect, which are magnified in (d) and (e), respectively with 1 nm scale bars.

Figure 1(a) illustrates the atomic structure of the 4Hb-TaS$_2$ crystal, which is composed of alternating 1T and 1H layers. A lattice of SoD clusters is observed on the 1T layer at both 4.7 K and room temperature. The atomic structure of these SoD clusters, overlaid in Fig. 1(b), shows the identical $\sqrt{13}a \times \sqrt{13}a$ periodicity at both temperatures. The clear CDW structure distinguishes the 1T layers from the 1H layers [Fig. S1]. Native CDW defects are identified in a larger STM image. The blue and red arrows in Fig. 1(c) indicate two representative types of CDW defects. The blue arrow denotes a Type 1 defect, which replaces a pristine SoD cluster with a "dark hole" at positive bias voltage of +200 mV [Fig. 1(d)]. Most dark holes exhibit the minimum height at off-centered positions and evolve into two off-centered clusters at negative bias voltages. In rare case, corresponding to 8 % of the observed Type 1 defects, the minimum height is located at the center of a dark hole, transforming into a single cluster at negative bias voltages [Fig. S2]. In contrast, a Type 2 defect preserves the SoD cluster but reduces the height in STM image [Fig. 1(e)]. Among Type 2 defects, the topographic contrast relative to pristine SoD clusters can vary, but no shape deformation is observed [Fig. S2(c-d)]. These topographic features suggest that Type 1 defects distort atomic structure while Type 2 defects primarily perturb the electronic configuration of SoD clusters.

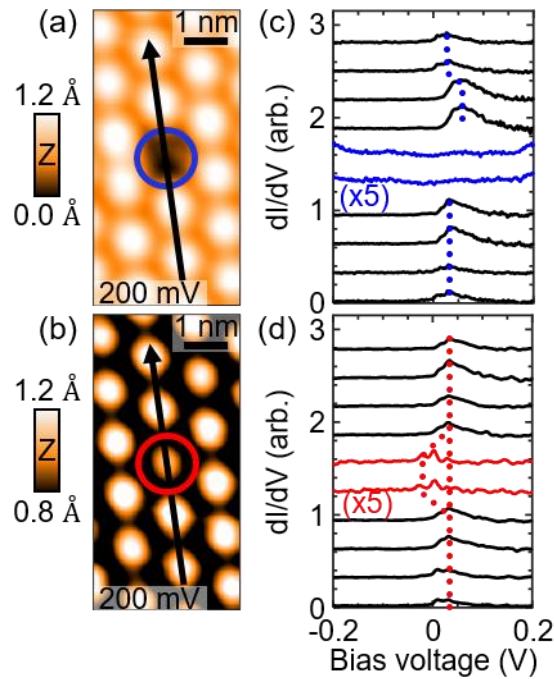

Fig. 2. (a, b) STM images of CDW defects and their neighboring SoD clusters. A blue circle in (a) denotes a Type 1 defect, while the red circle in (b) indicates a Type 2 defect. (c, d) d$I$/d$V$ spectra measured across the CDW defects along the black arrows in (a) and (b), respectively. (c) The blue spectra are measured on the Type 1 defect, while (d) the red spectra are measured on the Type 2 defects. Dotted lines serve as visual guides, tracking the transition of the flat band peaks across the CDW defects. All spectra are offset by equal intervals for clarity, and the d$I$/d$V$ spectra measured on the CDW defects are amplified 5 times for better visibility.

To evaluate the effect of CDW defects on the electronic structure, differential tunneling conductance (d$I$/d$V$) spectra were measured across each type of CDW defect. Figure 2(a) shows an asymmetrically shaped Type 1 defect, indicated by a blue circle, alongside neighboring SoD clusters. d$I$/d$V$ spectra, measured across the Type 1 defect, are displayed in Fig. 2(c). On pristine SoD clusters, far from the defect (top and bottom spectra), unoccupied flat bands manifest as a single peak at around 40 meV, consistent with previous reports [12,18]. However, near the Type 1 defect, the flat band peak shifts toward higher energy, as indicated by the dotted line in Fig. 2(c). Right on top of the Type 1 defect [blue spectra in Fig. 2(c)], the flat band peak is suppressed, and a new spectral peaks emerge [Fig. S3(a-d)]. These spectral features also appear around a symmetrically shaped Type 1 defect, but within a symmetric effective region. Only a single defect state emerged [Fig. S3], and the shifted flat band peaks are observed on both nearest-neighbor clusters [Fig. S4]. Closely related defect states and atomic configuration indicate that the lattice distortion dominates the electronic structure of Type 1 defects.

Next, d$I$/d$V$ spectra were measured across a Type 2 defect, denoted by a red circle in Fig. 2(c). The red dotted line in Fig. 2(d) indicates the transition of flat bands across the Type 2 defect. A distinct transition occurs on top of the Type 2 defect [red spectra in Fig. 2(d)], where the single flat band peak splits into three characteristic peaks around $E_F$. Previous DFT studies indicated that flat bands gradually split into lower and upper Hubbard bands as the energy of the flat bands approaches $E_F$ [21]. As a result, we can assign the occupied and unoccupied peaks to the lower Hubbard band (LHB) and upper Hubbard band (UHB), respectively. The energy difference between the two Hubbard bands, reflecting the on-site

Coulomb energy, is determined by the flat bands filling factor of the 1T layer. On the Type 2 defect, the Hubbard bands are separated by 50 meV. This smaller energy difference, compared to the 250 ~ 400 meV observed in 1T-TaS$_2$ crystal [22-24], shows a reduced on-site Coulomb energy reflecting the fractional occupancy of unpaired electrons. The electron doping at the Type 2 defect is further supported by the appearance of a zero-bias peak. A Fano line [Fig. S5(a)] fits well the zero-bias peak as previously reported Kondo-like resonance peak in 4Hb-TaS$_2$ [18], suggesting that Type 2 defects are likely electron modulated SoD clusters.

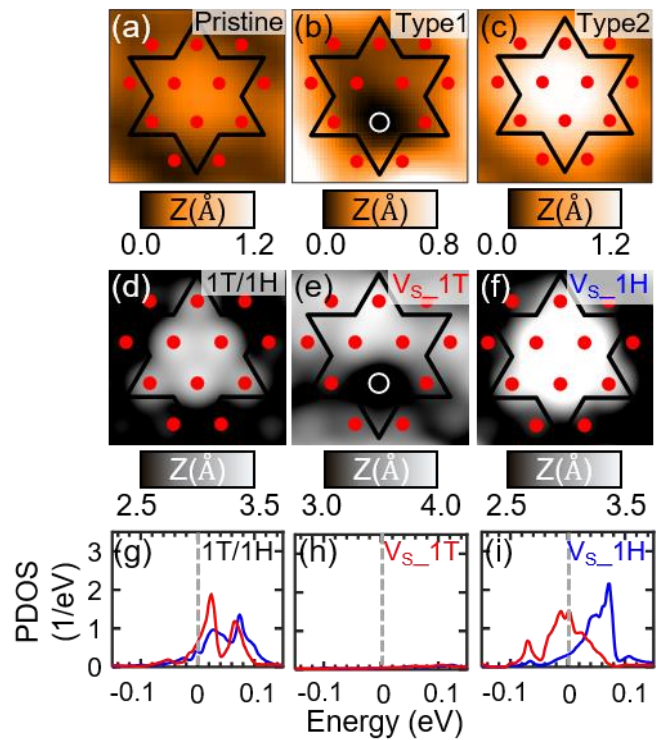

Fig.3 (a-c) STM images of (a) a pristine SoD cluster, (b) a Type 1 defect, and (c) a Type 2 defect. (d-f) Simulated STM images from three structural models; (d) a 1T/1H bilayer, (e) a 1T/1H bilayer with a $V_S\_1T$, and (f) a 1T/1H bilayer with a $V_S\_1H$. The bias voltage of 200 mV was used for both (b) and (e), while the other STM images were measured or simulated at a bias voltage of -200 mV. White circles in (b) and (e) denote the site of sulfur vacancies. (g-i) The projected density of states (PDOS) for the $d_{z^2}$ orbital from the three structural models. Red and blue lines correspond to spin-up and spin-down PDOS, respectively.

To understand the origin of CDW defects, we theoretically simulated STM images and electronic structures using first-principles calculations. For pristine SoD clusters, we selected a 1T/1H bilayer. The STM image of a pristine SoD cluster, shown in the magnified view in Fig. 3(a), reveals that three S-atoms at the center of the cluster appear higher than those at the periphery. Such topographic feature is well reproduced in the STM simulation of the 1T/1H bilayer structure [Fig. 3(d)]. The relaxed interlayer spacing of 5.97 Å is also in good agreement with the experimentally measured 5.9 Å [Fig. S1]. We have considered native atomic defects, including Ta vacancies, S vacancies, and anti-site substitutions, and found that S vacancies are associated with both types of CDW defects. The atomic structure of the SoD cluster, overlaid on the STM image in Fig. 3(b), highlights the minimum height at the S site in the Type 1 defect. This topographic feature is reproduced in the STM simulation when the 1T/1H bilayer structure contains a sulfur vacancy in the 1T layer ($V_S\_1T$) [Fig. 3(e)]. In contrast, a sulfur vacancy located in the 1H layer ($V_S\_1H$) exhibits an increased height [Fig. 3(f)] compared to the pristine 1T/1H bilayer [Fig. 3(d)], which matches the brighter SoD cluster seen in Type 2 defects [Fig. 3(c)] compared to pristine SoD clusters [Fig. 3(a)].

Next, using our models of CDW defects, we calculated band structures to investigate their effect on the flat bands of 4Hb-TaS$_2$ [Fig. S8]. Given the dominant contribution of Ta $5d_{z^2}$ orbital to the flat bands [10,25], we calculated the projected density of states (PDOS) of the $d_{z^2}$ orbital, as presented in Fig. 3(g-i). In the 1T/1H bilayer, the PDOS (g) display merged peaks above $E_F$, originating from flat bands, as confirmed in d$I$/d$V$ spectra of pristine SoD clusters. Introducing a $V_S\_1T$ in the bilayer dramatically decreases the PDOS (h) around $E_F$, supporting the locally suppressed flat band peak observed at Type 1 defects. In contrast, for a $V_S\_1H$, the flat bands split into a pair of Hubbard bands located above and below $E_F$ (i). Since anion vacancies are generally expected to act as electron dopants, a S vacancy would release electrons. However, in TaS$_2$, the effective range of the direct doping effect is confined to a single SoD cluster [26,27]. Hence, the observed hole-doping effect around Type 1 defects is likely the result of indirect influence from a S vacancy on surrounding SoD clusters, mediated by donated local charge, impurity potential, and atomic distortions. On the other hand, a $V_S\_1H$ would donate electrons to the 1H layer and suppress the charge transfer from the 1T layer to the 1H layer, thereby effectively leading to the observed electron doping in the 1T layer.

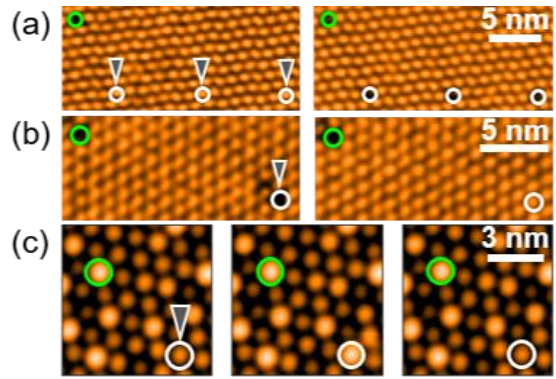

Fig.4. (a) STM images before (left panel) and after (right panel) creating three Type 1 defects. White circles highlight the transformation of pristine SoD clusters (left panel) into Type 1 defects (right panel). (b) STM images before (left panel) and after (right panel) erasing a Type 1 defect. The white circle highlights the transition from a Type 1 defect (left panel) back into a pristine SoD cluster (right panel). (c) STM images before (left panel) and after (middle panel) creating a Type 2 defect. White circles highlight the transformation of a pristine SoD cluster (left panel) into a Type 2 defect (middle panel), and then the spontaneous recovery to a pristine SoD cluster (right panel). Pre-existing defects are denoted by green circles for position reference. The grey triangle indicates the STM-tip position when a bias pulse or high tunneling current is applied. STM images were measured at (a-b) the bias voltage of 200 mV in room temperature and (c) -200 mV in 4.7 K.

Since it is known that the STM tip can manipulate atomic defects on surfaces [19,20,28,29], we applied this technique to control individual CDW defects to rationally modulate the flat band. By applying a bias pulse and a high tunneling current, we achieved the bi-directional manipulation of Type 1 defects at room temperature. A pristine SoD cluster was converted into a Type 1 defect by applying a 2.7 V bias pulse [Fig. 4(a)]. The created Type 1 defect remained stable on the surface, allowing us to generate multiple defects with uniform spacing, as shown in lower panel of Fig. 4(a). Given the lower formation energy of S vacancies compared to Ta vacancies in 4Hb-$TaS_2$ [Fig. S10], the bias pulse creates a Type 1 defect by inducing a new $V_S\_1T$, similar to the vacancy formation induced by electron beams in 1T-

TaS$_2$ crystal [30]. Conversely, Type 1 defects were erased by bringing the STM tip closer and increasing the tunneling current from 300 pA to 1.4 nA [Fig. 4(b)]. As the STM tip was pre-cleaned on a Cu(111) single crystal, it is unlikely that the S atom filling the vacancy originated from the tip itself. Instead, it likely diffused from the underlying 1H layer. Approaching STM tip closer to the surface may lower the energy barrier for S atom diffusion between the 1T and 1H layers, a phenomenon both theoretically expected and experimentally observed in layered materials [19,20,31,32]. Measurements of d$I$/d$V$ spectra after each manipulation step confirmed the creation of defect states after the writing process and the restoration of the flat band peak after the erasing process [Fig. S6]. Identical STM/S features indicate that the created Type 1 defects match the pre-existing ones, consistently indicating that V$_S$_1T can host Type 1 defects. At the low temperature of 4.7 K, interactions between the STM tip and Type 1 defects are not observed, but applying a bias pulse transformed a pristine SoD cluster into a Type 2 defect [Fig.4 (c)]. The created Type 2 defect spontaneously reverted to a pristine SoD cluster at 4.7 K. Since the temperature and scanning conditions were held constant—both of which have previously been identified as possible driving forces for transitions between Type 2 defects and pristine SoD clusters [18]—our observation suggests a spontaneous transition of charge density defects in 4Hb-TaS$_2$.

Our study demonstrates that electron correlation in the 1T layer can be controlled by manipulating CDW defects, an important capability that has been lacking in efforts to understand its influence on superconductivity in the adjacent 1H layer. Type 1 defects, which disturb flat band electrons, are expected to suppress spin fluctuations in a doped Mott insulator—a mechanism that has been suggested as a possible origin of chiral superconductivity [13]. In contrast, Type 2 defects suppress interlayer charge transfer and thereby break inversion symmetry in the superconducting 1H layer, a proposed mechanism for the emergence of topologically protected boundary modes [15]. Investigating how superconducting properties respond to controlled CDW defect will offer a systematic method to uncover the interplay between correlated electrons and superconductivity to clarify the mechanism of non-trivial superconductivity.

In summary, we investigated the atomic and electronic structures of CDW defects in 4Hb-TaS$_2$ using STM/S measurements and first principles calculations. Our study identified two types of CDW defects:

Type 1 and Type 2. $V_S\_1T$ locally distorts the SoD cluster, leading to the spectral features of Type 1 defects, such as suppression of flat bands, emergence of defect states, and hole-doping in neighboring clusters. In contrast, $V_S\_1H$ locally reduce interlayer coupling that causes Type 2 defects. Furthermore, we experimentally demonstrated the bi-directional manipulation of individual defects either by controlling S vacancies or inducing metastable states using the STM tip. This reversible and selective manipulation allows precise tuning of CDW via interaction with atomic lattice distortion or interlayer coupling. Our ability to manipulate individual CDW defects provides a novel avenue for tuning strongly correlated states with atomic precision.

**Acknowledgments**

This work was conducted at the Center for Nanophase Materials Sciences (CNMS), which is a U.S. Department of Energy, Office of Science User Facility. The crystal growth (H.D.Z) was supported by the U.S. Department of Energy (DOE) under Grant No. DE-SC0020254. S.M. acknowledges the startup fund from the University of South Carolina and an Advanced Support for Innovative Research Excellence (ASPIRE) grant from the Office of the Vice President for Research at the University of South Carolina. This work used the Expanse supercomputer at the San Diego Supercomputer Center through allocation PHY230093 from the Advanced Cyberinfrastructure Coordination Ecosystem: Services & Support (ACCESS) program, which is supported by National Science Foundation Grants No. 2138259, No. 2138286, No. 2138307, No. 2137603, and No. 2138296. This research also used resources of the National Energy Research Scientific Computing Center (NERSC), a U.S. Department of Energy Office of Science User Facility operated under Contract No. DE-AC02-05CH11231.